\documentclass[11pt]{article}

\usepackage{graphicx}
\usepackage{amsmath}
\usepackage{latexsym}
\usepackage{amssymb}

\DeclareSymbolFont{AMSb}{U}{msb}{m}{n}
\DeclareMathSymbol{\N}{\mathbin}{AMSb}{"4E}
\DeclareMathSymbol{\Z}{\mathbin}{AMSb}{"5A}
\DeclareMathSymbol{\R}{\mathbin}{AMSb}{"52}
\DeclareMathSymbol{\Q}{\mathbin}{AMSb}{"51}
\DeclareMathSymbol{\I}{\mathbin}{AMSb}{"49}
\DeclareMathSymbol{\C}{\mathbin}{AMSb}{"43}

\begin{document}

\title{Stress tensor for massive fields on flat spaces of spatial topology $\R^2\times{S^1}$}
\author{Paul Langlois\footnote{Electronic Address: pmxppl@nottingham.ac.uk} 
\\ \textit{School of Mathematical Sciences, University of Nottingham} 
\\ 
\textit{Nottingham NG7 2RD, UK}}
\date{\today}
\maketitle

\begin{abstract}

 We calculate the expectation values of the energy-momentum tensor $T_{{\mu}{\nu}}$ 
 for massive scalar and spinor fields, in the Minkowski-like vacuum states 
 on the two flat spaces which are quotients of Minkowski space under the
 discrete isometries $(t,x,y,z)\mapsto(t,x,y,z+2a)$ and $(t,x,y,z)\mapsto(t,-x,-y,z+a)$.
 The results on the first space confirm the literature. 
 The results on the second space are new. We note some
 qualitative differences between the massless and massive fields
 in the limits of large $a$ and large $x^2+y^2$.

\end{abstract}

\section{Introduction}

We present the expectation values of the energy-momentum 
tensor, in the Minkowski-like vacuum, for massive scalar and spinor fields in the 
flat but topologically non-trivial spacetimes $M_0$ and $M_-$. These
are quotients of Minkowski space under the groups generated 
by the discrete isometries
$J_0:(t,x,y,z)\mapsto(t,x,y,z+2a)$ and $J_-:(t,x,y,z)\mapsto(t,-x,-y,z+a)$,
respectively. For details of 
these spacetimes and for the calculations in the massless case see 
\cite{lm:geon,pl:langlois}~\footnote{The expectation values for the massless scalar field 
on $M_0$ appear in a number of places in the literature \cite{lm:geon,bd:em,dw:dewitthi}. 
Those for $2$-component spinors are 
also found in \cite{bd:em,dw:dewitthi}.}. The expectation values for massive scalar and spinor
fields on 
$M_0$ are in the literature \cite{th:scalar,an:spinor}. 
The results for the massive fields on $M_-$ are new.
It is seen that in all cases the values decay exponentially in the
limit of large mass, and the leading order corrections for small mass
are $O(m^2)$. Further we note some qualitative differences between the
behaviour of the massless and massive field values in the limits of large $a$ and
large $x^2+y^2$.
Interest in this problem arises from the role of 
$M_0$ and $M_-$ in modelling, via accelerated observers on 
flat spacetimes, the Hawking(-Unruh) effect on respectively the Kruskal 
manifold and the $\mathbb{RP}^3$ geon \cite{lm:geon}.
As an aside we also present the expectation values of the stress
tensor, in the Minkowski-like
vacuum, for a massive scalar field on Minkowski space with an infinite
straight plane boundary.

We work throughout in natural units $\hbar=c=1$ and with metric 
signature $(+,-,-,-)$.

\section{The massive scalar field}

The case of massive scalar fields in multiply connected flat spacetimes was considered
by Tanaka and Hiscock in \cite{th:scalar}. In particular they consider 
$\langle{0}|T_{\mu\nu}|0\rangle$ on flat spacetimes 
with topology $\R^3\times{S^1}$ (which is denoted $M_0$ here), $\R^2\times{T^2}$ and $\R^1\times{T^3}$.
It is seen in \cite{th:scalar} that the magnitude of the energy density decreases with an increasing 
field mass.
Here we will reproduce the result on $M_0$ and present the expectation values on $M_-$.

The energy-momentum tensor for the massive scalar field in a general four-dimensional
curved spacetime in our conventions is \cite{bd:book}
\begin{eqnarray}
T_{\mu\nu} & = &
(1-2\xi)\phi_{;\mu}\phi_{;\nu}
+
(2\xi-\frac{1}{2})g_{\mu\nu}g^{\rho\sigma}\phi_{;\rho}\phi_{;\sigma}
-
2\xi\phi_{;\mu\nu}\phi
\nonumber
\\
&  & +
\frac{1}{2}\xi{g_{\mu\nu}}\phi\Box\phi
-\xi
\left[
R_{\mu\nu}
-\frac{1}{2}Rg_{\mu\nu}+\frac{3}{2}\xi{R}g_{\mu\nu}\right]
\phi^2
\nonumber
\\
& & + \frac{1}{2}\left[1-3\xi\right]m^2g_{\mu\nu}\phi^2
 \ , 
\end{eqnarray}
which, with the help of the field equation $[\Box+m^2+\xi{R}]\phi=0$, may be
written on a flat spacetime as
\begin{eqnarray}
T_{\mu\nu} & = &
(1-2\xi)\phi_{;\mu}\phi_{;\nu}
+
(2\xi-\frac{1}{2})g_{\mu\nu}g^{\rho\sigma}\phi_{;\rho}\phi_{;\sigma}
-
2\xi\phi_{;\mu\nu}\phi
\nonumber
\\
&  & +
2\xi{g_{\mu\nu}}\phi\Box\phi
+\frac{1}{2}m^2g_{\mu\nu}\phi^2
 \ ,
\end{eqnarray}
where $\xi$ gives the coupling to the gravitational field ($\xi=0$ for minimal
coupling and $\xi=1/6$ for conformal coupling). By the point splitting 
technique \cite{ch:pointsplit}, 
where we split the points in the above quadratic expressions and take the coincidence
limit at the end, this may be written as 
\begin{eqnarray}
T_{\mu\nu} & = &
\frac{1}{2}
\lim_{x'\rightarrow{x}}
\left[
(1-2\xi)\nabla_\mu\nabla_{\nu'}
+
(2\xi-\frac{1}{2})g_{\mu\nu}\nabla_\alpha\nabla^{\alpha'}
-
2\xi\nabla_\mu\nabla_{\nu}
\right.
\nonumber
\\
&  & +
\left.
2\xi{g_{\mu\nu}}\nabla_\alpha\nabla^{\alpha}
+\frac{1}{2}m^2g_{\mu\nu}
\right]
\{\phi(x),\phi(x')\}
 \ ,
\end{eqnarray}
where $\{a,b\}$ is an anticommutator.

The spacetimes we consider here are $M_0$ and $M_-$. These are built as quotients
of Minkowski space under certain discrete isometry groups. Both spacetimes admit a
global timelike Killing vector $\partial_t$ and a vacuum state built from it (which 
we denote by $|0\rangle$ in all cases, the particular vacuum being considered being
that relevant for the spacetime under consideration). This 
vacuum state is that induced by the usual vacuum on Minkowski space. The expectation value 
of $T_{\mu\nu}$ in this vacuum state is then
\begin{eqnarray}
\langle{0}|T_{\mu\nu}|0\rangle & = &
\frac{1}{2}
\lim_{x'\rightarrow{x}}
\left[
(1-2\xi)\nabla_\mu\nabla_{\nu'}
+
(2\xi-\frac{1}{2})g_{\mu\nu}\nabla_\alpha\nabla^{\alpha'}
-
2\xi\nabla_\mu\nabla_{\nu}
\right.
\nonumber
\\
\label{eqn:emtensor}
&  & +
\left.
2\xi{g_{\mu\nu}}\nabla_\alpha\nabla^{\alpha}
+\frac{1}{2}m^2g_{\mu\nu}
\right]
G^{(1)}(x,x')
 \ ,
\end{eqnarray}
where $G^{(1)}(x,x')=\langle{0}|\{\phi(x),\phi(x')\}|0\rangle=G^+(x,x')+G^-(x,x')$ is the scalar 
Hadamard function which on Minkowski space is given by (see appendix C of \cite{bj:relfields})~\footnote{Note 
the sign difference in the second term from that in Fulling's book \cite{fu:Fulling} p85
which is the reference used by Tanaka and Hiscock \cite{th:scalar}. Further 
note the typographical error in \cite{th:scalar}, where the $Y$ Bessel has been replaced by the $I$ Bessel function. 
This does not affect the results in \cite{th:scalar}.}
\begin{eqnarray}
G^{(1)}(x,x') & = & \frac{m}{4\pi{((t-t')^2-R^2)^{1/2}}}Y_1(m\sqrt{(t-t')^2-R^2})\Theta((t-t')^2-R^2)
\nonumber
\\
\nonumber
& & +\frac{m}{2\pi^2(R^2-(t-t')^2)^{1/2}}K_1(m\sqrt{R^2\!-\!(t-t')^2})\Theta(R^2\!-\!(t-t')^2) \ , 
\\
\label{eqn:minkhad}
\end{eqnarray}
where $Y,K$ are Bessel functions and $R^2=(x-x')^2+(y-y')^2+(z-z')^2$. 
The Hadamard function on the quotient spaces $M_0$ and $M_-$ may be found by the method
of images \cite{bD:auto,bd:em}, as
\begin{eqnarray}
\label{eqn:m0had}
G^{(1)}_{M_0}(x,x') & = & \sum_{n\in{\Z}}\eta^nG^{(1)}(x,J_0^nx')
\\
\label{eqn:m-had}
G^{(1)}_{M_-}(x,x') & = & \sum_{n\in{\Z}}\eta^nG^{(1)}(x,J_-^nx') \ , 
\end{eqnarray}
where $J_0:(t,x,y,z)\mapsto(t,x,y,z+2a)$, $J_-:(t,x,y,z)\mapsto(t,-x,-y,z+a)$ and
$\eta=1, (-1)$ labels standard (twisted) fields, respectively.
The calculation of $\langle{0}|T_{\mu\nu}|0\rangle$ on these spaces is now reduced to that
of finding derivatives of these Hadamard functions and applying (\ref{eqn:emtensor}) to (\ref{eqn:m0had})
and (\ref{eqn:m-had}). Renormalization is achieved as usual 
on these flat spaces \cite{bd:book} by dropping the divergent Minkowski contribution
coming from the $n=0$ terms in the above sums. Next we present the results.

\subsection{$M_0$}

For $M_0$, in the coincidence limit $G^{(1)}_{M_0}(x,x')$ becomes
a function only of $(2na)^2$ which is positive. This means that only the
$K$ Bessel term in (\ref{eqn:minkhad}) contributes. Further we note
that $\lim_{x'\rightarrow{x}}\partial_{x^\mu}G^{(1)}_{M_0}(x,x')
=-\lim_{x'\rightarrow{x}}\partial_{{x'}^\mu}G^{(1)}_{M_0}(x,x')$.
These observations simplify the calculations somewhat. The result is
\begin{eqnarray}
\langle{0}|T_{tt}|0\rangle_{M_0}=-\langle{0}|T_{xx}|0\rangle_{M_0}=-\langle{0}|T_{yy}|0\rangle_{M_0}
=
-\sum_{n=1}^\infty
\eta^n
\frac{m^2}{2\pi^2(2na)^2}
K_2(2mna)
\nonumber
\\
\label{eqn:scalarM0}
\langle{0}|T_{zz}|0\rangle_{M_0}
=
\sum_{n=1}^\infty
\eta^n
\left[\frac{m^2}{2\pi^2(2na)^2}
K_2(2mna)
-
\frac{m^3}{2\pi^2(2na)}
K_3(2mna)\right] \ , 
\end{eqnarray}
with all other components vanishing. The results agree with \cite{th:scalar}.
The massless limit may be easily checked by noting that near $z=0$
\begin{equation}
K_\nu(z)
=
\frac{\Gamma(\nu)}
{2}
(\frac{z}{2})^{-\nu} \ ,
\end{equation}
when the real part of $\nu$ is positive \cite{gr:inttables}. These agree with
\cite{lm:geon,dw:dewitthi} and also with
\cite{bd:em}, the extra factor of $4$ in \cite{bd:em} arising because the
authors consider a multiplet of two complex massless scalar fields.
The leading corrections for small mass are of the order $O(m^2)$. 
$\langle{0}|T_{\mu\nu}|0\rangle\rightarrow{0}$ exponentially in the
large mass and large $a$ limits, in contrast to the massless case 
where $\langle{0}|T_{\mu\nu}|0\rangle$ vanishes as $O(a^{-4})$.

\subsection{$M_-$}

For $M_-$ in the coincidence limit $G^{(1)}_{M_-}(x,x')$ becomes
a function only of $(x-(-1)^nx)^2+(y-(-1)^ny)^2+(na)^2$ which is positive. Again
only the $K$ Bessel term in (\ref{eqn:minkhad}) contributes.
Further we note the following
\begin{eqnarray}
\lim_{x'\rightarrow{x}}\partial_{\mu}G^{(1)}_{M_-}(x,x')
& = & -\lim_{x'\rightarrow{x}}\partial_{{\mu'}}G^{(1)}_{M_-}(x,x')
\;\; \textrm{for}\; \mu\in\{t,z\}
\nonumber
\\
\lim_{x'\rightarrow{x}}\partial_{\mu}G^{(1)}_{M_-}(x,x')
& = & -(+)\lim_{x'\rightarrow{x}}\partial_{{\mu'}}G^{(1)}_{M_-}(x,x')
\;\;
\textrm{when $n$ even (odd) and}\; \mu\in\{x,y\}.
\nonumber
\\
\label{eqn:m-relation}
&   & 
\end{eqnarray}
As $G^{(1)}_{M_-}(x,x')=\sum_{n\in{\Z}}\rho^nG^{(1)}(x,J_-^nx')$, where 
$\rho=+1, (-1)$ labels untwisted (twisted) fields,
the result here may be split into two parts where the part coming from the
even terms in the above sum leads to the same expectation values
as on $M_0$ for untwisted fields. Therefore we write
$\langle{0}|T_{\mu\nu}|0\rangle_{M_-}=\langle{0}|T_{\mu\nu}|0\rangle_{M_0(\eta=1)}+\rho
\langle{0}|T_{\mu\nu}|0\rangle_{odd}$ where we find
\begin{eqnarray}
\langle{0}|T_{tt}|0\rangle_{odd}
& = & 
\sum_{n\in{\Z}}
\left[
-(4\xi-1)\frac{m^3}{4\pi^2\sigma_n}
K_3(m\sigma_n)
\left[
1-\frac{(2na+a)^2}{\sigma_n^2}
\right]
\right.
\nonumber
\\
&   &\;\;\;\;\;\;\;+\left.
\left(2\xi-\frac{3}{4}\right)\frac{m^2}{\pi^2\sigma_n^2}
K_2(m\sigma_n)\right]
\nonumber
\\
\langle{0}|T_{xx}|0\rangle_{odd}
& = & 
\sum_{n\in{\Z}}
\left[
(4\xi-1)\frac{m^3y^2}{\pi^2\sigma_n^3}
K_3(m\sigma_n)
-
\left(2\xi-\frac{1}{2}\right)\frac{m^2}{2\pi^2\sigma_n^2}
K_2(m\sigma_n)\right]
\nonumber
\\
\langle{0}|T_{yy}|0\rangle_{odd}
& = & 
\sum_{n\in{\Z}}
\left[
(4\xi-1)\frac{m^3x^2}{\pi^2\sigma_n^3}
K_3(m\sigma_n)
-
\left(2\xi-\frac{1}{2}\right)\frac{m^2}{2\pi^2\sigma_n^2}
K_2(m\sigma_n)\right]
\nonumber
\\
\langle{0}|T_{zz}|0\rangle_{odd}
& = & 
\sum_{n\in{\Z}}
\left[
\frac{m^3}{4\pi^2\sigma_n}
K_3(m\sigma_n)
\left[
(4\xi-1)-\frac{4\xi(2na+a)^2}{\sigma_n^2}
\right]
\right.
\nonumber
\\
&   &\;\;\;\;\;\;\;-\left.
\left(2\xi-\frac{3}{4}\right)\frac{m^2}{\pi^2\sigma_n^2}
K_2(m\sigma_n)\right]
\nonumber
\\
\langle{0}|T_{xy}|0\rangle_{odd}
& = & 
\sum_{n\in{\Z}}
(1-4\xi)\frac{m^3xy}{\pi^2\sigma_n^3}
K_3(m\sigma_n) \ ,
\end{eqnarray}
where $\sigma_n=((2x)^2+(2y)^2+(2na+a)^2)^{1/2}$ and the
sum is over all $n$ including $n=0$. Other components vanish. Again
it is a simple matter to check the massless limit.
The results agree with those of \cite{lm:geon,bd:em} in this limit.
The leading correction for small mass is $O(m^2)$, and $\langle{0}|T_{\mu\nu}|0\rangle$
vanishes exponentially in the large mass and, for non-zero mass, the
large $a$ limits. The difference between $\langle{0}|T_{\mu\nu}|0\rangle$ on $M_-$
and $M_0$ vanishes exponentially as $r^2:=x^2+y^2\rightarrow\infty$.
This behaviour is qualitatively different to the
massless case where the difference vanishes as $O(r^{-3})$.
% what about the massless limit for large a? exponential or a^-4?
The result in the massive case is to our knowledge new.

\section{The massive Dirac field}

In this section we repeat the above calculations for the massive
Dirac field. The result on $M_0$ was given recently in \cite{an:spinor}
where it is shown that, as for the scalar field, the magnitude of energy density
decreases with increasing field mass.
We shall present the results for $M_0$ and $M_-$ and comment on 
various limits.

The energy-momentum tensor for the massive Dirac field is \cite{bd:book}
\begin{equation}
T_{\mu\nu}
=
\frac{i}{2}
[
\bar{\psi}\gamma_{(\mu}\nabla_{\nu)}\psi-\nabla_{(\mu}\bar{\psi}\gamma_{\nu)}\psi
] \ ,
\end{equation}
where $A_{(\mu}B_{\nu)}=1/2(A_\mu{B_\nu}+A_\nu{B_\mu})$. This may
be written as
\begin{equation}
T_{\mu\nu}
=
\frac{i}{4}
\mathrm{Tr}
(
\gamma_{(\mu}[\nabla_{\nu)}\psi,\bar{\psi}]-\gamma_{(\mu}[\psi,\nabla_{\nu)}\bar{\psi}]
) \ .
\end{equation}
Further we use the point splitting technique \cite{ch:pointsplit}
to write this as
\begin{eqnarray}
T_{\mu\nu}
& = & 
\frac{i}{8}
\lim_{x'\rightarrow{x}}
\mathrm{Tr}
(
\gamma_{(\mu}[\nabla_{\nu')}\psi(x'),\bar{\psi}(x)]+
\gamma_{(\mu}[\nabla_{\nu)}\psi(x),\bar{\psi}(x')]
\nonumber
\\
&   & 
-\gamma_{(\mu}[\psi(x),\nabla_{\nu')}\bar{\psi}(x')]
-\gamma_{(\mu}[\psi(x'),\nabla_{\nu)}\bar{\psi}(x)]
) \ .
\end{eqnarray}
The expectation value of $T_{\mu\nu}$ in the vacuum state
$|0\rangle$ may now be expressed in terms of the spinor 
Hadamard function $S^{(1)}_{\alpha\beta}(x,x')=\langle{0}|[\psi_\alpha(x),\bar{\psi}_\beta(x')]|0\rangle$
\begin{eqnarray}
\langle{0}|T_{\mu\nu}|0\rangle
& = & 
\frac{i}{8}
\lim_{x'\rightarrow{x}}
\mathrm{Tr}
(
\gamma_{(\mu}\nabla_{\nu')}S^{(1)}(x',x)+
\gamma_{(\mu}\nabla_{\nu)}S^{(1)}(x,x')
\nonumber
\\
&   & 
-\gamma_{(\mu}\nabla_{\nu')}S^{(1)}(x,x')
-\gamma_{(\mu}\nabla_{\nu)}S^{(1)}(x',x)
) \ .
\end{eqnarray}
Further the spinor Hadamard function may be expressed in terms of the
scalar one $S^{(1)}(x,x')=-(i\gamma^\rho\nabla_\rho+m)G^{(1)}(x,x')$,
where $G^{(1)}(x,x')$ is given by (\ref{eqn:minkhad}) in Minkowski space, 
and we note that
\begin{equation}
\lim_{x'\rightarrow{x}}\nabla_\mu{S^{(1)}(x',x)}
=\lim_{x'\rightarrow{x}}\nabla_{\mu'}{S^{(1)}(x,x')}
\end{equation}
and so
\begin{eqnarray}
\langle{0}|T_{\mu\nu}|0\rangle
& = & 
\frac{i}{4}
\lim_{x'\rightarrow{x}}
\mathrm{Tr}
\gamma_{(\mu}
(\nabla_{\nu)}S^{(1)}(x,x')
-\nabla_{\nu')}S^{(1)}(x,x'))
\nonumber
\\
& = & 
\frac{1}{8}
\lim_{x'\rightarrow{x}}
\mathrm{Tr}
[(\gamma_{\mu}\nabla_{\nu}+\gamma_{\nu}\nabla_{\mu})
-
(\gamma_{\mu}\nabla_{\nu'}+\gamma_{\nu}\nabla_{\mu'})]
\gamma^\rho\nabla_\rho
G^{(1)}(x,x') \ .
\nonumber
\\
\label{eqn:spinexpect}
\end{eqnarray}
Note that this expression is twice that of (10) of reference \cite{an:spinor}
where the Majorana spinors are considered.

We are concerned  with these expectation values on quotient spaces of Minkowski 
space. As with the scalar field the Hadamard function may be found here
by the method of images, but extra care must be taken with the local Lorentz
frames with respect to which the spinors are expressed. We shall work throughout 
with a vierbein aligned along the usual Minkowski coordinate axes. This has the
advantage of making covariant and partial derivatives coincide. As this 
vierbein is invariant under $J_0$ the 
calculation on $M_0$ is then reduced to a straightforward calculation of these
derivatives of the Hadamard function and applying (\ref{eqn:spinexpect}).
The Minkowski vierbein however is not invariant under $J_-$ and 
more care must be taken on $M_-$.

\subsection{$M_0$}

Recall that on $M_0$
\begin{equation}
G^{(1)}_{M_0}(x,x')
=
\sum_{n\in{\Z}}
\eta^n
G^{(1)}(x,J_0^nx') \ , 
\end{equation}
where renormalization is performed again by simply dropping the Minkowski $n=0$
term in the sum.
Here again we note that in the coincidence limit $G^{(1)}_{M_0}(x,x')$ becomes
a function only of $(2na)^2$ which is positive and so only the
$K$ Bessel term in (\ref{eqn:minkhad}) contributes. Also
$\lim_{x'\rightarrow{x}}\partial_{x^\mu}G^{(1)}_{M_0}(x,x')
=-\lim_{x'\rightarrow{x}}\partial_{{x'}^\mu}G^{(1)}_{M_0}(x,x')$.
With these observations (\ref{eqn:spinexpect}) reduces to
\begin{equation}
\langle{0}|T_{\mu\nu}|0\rangle
=
\lim_{x'\rightarrow{x}}
\nabla_\mu\nabla_\nu
G^{(1)}_{M_0}(x,x') \ ,
\end{equation}
and we find for the non-zero expectation values
\begin{eqnarray}
\langle{0}|T_{tt}|0\rangle_{M_0}=-\langle{0}|T_{xx}|0\rangle_{M_0}=-\langle{0}|T_{yy}|0\rangle_{M_0}
=
\sum_{n=1}^\infty
\eta^n
\frac{2m^2}{\pi^2(2na)^2}
K_2(2mna)
\nonumber
\\
\langle{0}|T_{zz}|0\rangle_{M_0}
=
\sum_{n=1}^\infty
\eta^n\left[
-\frac{2m^2}{\pi^2(2na)^2}
K_2(2mna)
+
\frac{2m^3}{\pi^2(2na)}
K_3(2mna)
\right] \ ,
\end{eqnarray}
where $\eta=+1, (-1)$ labels periodic (twisted) spinors with respect to the standard Minkwoski 
vierbein (that is, $\eta$ labels the two possible spin structures on $M_0$ \cite{pl:langlois}).
The $\eta=-1$ spin structure is energetically preferred.
The results are $-4$ times those of the massive scalar field (\ref{eqn:scalarM0}). The factor of $-1$ is 
due to different statistics while the factor of $4$ is due to degrees 
of freedom. 
The massless limit agrees
with twice the massless two-component expectation values 
\cite{pl:langlois,bd:em,dw:dewitthi} as expected.

\subsection{$M_-$}

The standard Minkowski vierbein is not invariant under $J_-$. 
In an invariant vierbein the spinor Hadamard function on $M_-$
would be
given directly by the method of images, that is
\begin{equation}
S^{(1)}_{M_-}(x,x')
=
\sum_{n\in{\Z}}
\rho^n
S^{(1)}(x,J_-^nx') \ ,
\end{equation}
with $\rho=+1, (-1)$,
however as we choose to work in the Minkowski vierbein the image
expression is different. One vierbein which is invariant under $J_-$ 
is one which rotates by $\pi$ 
in the $x-y$-plane as $z\rightarrow{z+a}$. The transformation from this vierbein to
the standard Minkowski one is clearly the corresponding rotation by $-\pi$. The
associated transformation of the spinor Hadamard function is
\begin{equation}
S^{(1)}_{SM}(x,x')
=
e^{\frac{\pi\gamma^1\gamma^2}{2a}z}
S^{(1)}_{RM}(x,x')
e^{-\frac{\pi\gamma^1\gamma^2}{2a}z'} \ , 
\end{equation}
the $R (S)$ subscript denotes the rotating (standard) vierbein respectively.
Therefore on $M_-$ in the standard vierbein
\begin{equation}
S^{(1)}_{SM_-}(x,x')
=
\sum_{n\in{\Z}}
\rho^n
S^{(1)}_{SM_-}(x,J_-^nx')
e^{\frac{n\pi\gamma^1\gamma^2}{2}} \ .
\end{equation}
In terms of the scalar Hadamard function this translates
to using the following expression in (\ref{eqn:spinexpect}) 
\begin{equation}
\label{eqn:hadexpr}
G^{(1)}_{M_-}(x,x')
=
\sum_{n\in{\Z}}
\rho^n
G^{(1)}(x,J_-^nx')
e^{\frac{n\pi\gamma^1\gamma^2}{2}} \ ,
\end{equation}
where the $n=0$ term is dropped. Here $\rho=+1, (-1)$ labels periodic (antiperiodic)
spinors with respect to the vierbein which rotates by $\pi$ as $z\rightarrow{z+a}$. Thus,
$\rho$ labels the two inequivalent spin structures on $M_-$ \cite{pl:langlois}.
In the coincidence limit $G^{(1)}_{M_-}(x,x')$ becomes
a function only of $(x-(-1)^nx)^2+(y-(-1)^ny)^2+(na)^2$ which is positive so that again
only the $K$ Bessel term in (\ref{eqn:minkhad}) contributes. Further we note again
the relations (\ref{eqn:m-relation}). We now therefore just apply 
(\ref{eqn:spinexpect}) to (\ref{eqn:hadexpr}).
The calculation is made easier by splitting the sum in (\ref{eqn:hadexpr})
into odd and even terms. The even terms lead to the expectation values on $M_0$ in the
twisted spin structure there and $\langle{0}|T_{\mu\nu}|0\rangle_{M_-}=
\langle{0}|T_{\mu\nu}|0\rangle_{M_0(\eta=-1)}+\rho\langle{0}|T_{\mu\nu}|0\rangle_{odd}$
with
\begin{eqnarray}
\langle{0}|T_{zx}|0\rangle_{odd}
& = & 
-\sum_{n\in{\Z}}
(-1)^n\frac{m^3y(2na+a)}{\pi^2\sigma_n^3}
K_3(m\sigma_n)
\nonumber
\\
\langle{0}|T_{zy}|0\rangle_{odd}
& = & 
\sum_{n\in{\Z}}
(-1)^n\frac{m^3x(2na+a)}{\pi^2\sigma_n^3}
K_3(m\sigma_n) \ ,
\end{eqnarray}
where $\sigma_n=((2x)^2+(2y)^2+(2na+a)^2)^{1/2}$.

We see that the spinor expectation values on $M_-$ are not
$-4$ times those of the scalar field. In particular it is interesting to note that
for the scalar field the only non-zero cross term is $\langle{0}|T_{xy}|0\rangle$
while for the spinor field this term is $0$ and $\langle{0}|T_{xz}|0\rangle$
and $\langle{0}|T_{yz}|0\rangle$ are non-zero. Also note that these
two terms for the spinor field change sign under a change of spin structure.
In the massless limit the results agree with twice those of the
massless two-component spinor results found in \cite{pl:langlois} where it is
shown that the expectation values for right- and left-handed 
two-component spinors are the same (see also \cite{bd:em}). The 
leading order corrections for small mass are $O(m^2)$ and 
$\langle{0}|T_{\mu\nu}|0\rangle$ vanishes exponentially in the
limits of large mass and large $a$.
The difference between $\langle{0}|T_{\mu\nu}|0\rangle$ on
$M_-$ and on $M_0$ vanishes at $r^2:=x^2+y^2=0$ and
vanishes exponentially as $r\rightarrow\infty$.

\section{Minkowski space with infinite plane boundary}

In this section we consider briefly a massive
scalar field in four-dimensional Minkowski space with an infinite plane boundary
at $x=0$. While the stress-energy for a massless field is well known
(see e.g \cite{bd:book,d:dewitt}), the massive results to our knowledge 
are new. 
There is some similarity with
$M_-$ as both spaces may be considered as quotients of Minkowski space
with the quotient group including a reflection in $x$ about $x=0$.

Again here the scalar Hadamard function is given by the method of 
images,
\begin{equation}
G^+_B(x,x')=G^+_M(x,x')+\eta{G^+_M(x,J_Bx')} \ ,
\end{equation}
where $J_B:(t,x,y,z)\mapsto(t,-x,y,z)$ and $\eta=+1, (-1)$ labels
Neumann (Dirichlet) boundary conditions on the plate. 
The first term leads to the expectation values 
$\langle{0}|T_{\mu\nu}|0\rangle$ on Minkowski space
which are divergent and are dropped by the usual 
renormalization procedure. Therefore from the second term
we get
\begin{eqnarray}
\langle{0}|T_{tt}|0\rangle & = & -\langle{0}|T_{yy}|0\rangle=-\langle{0}|T_{zz}|0\rangle
\nonumber
\\
& = & \eta\left[
\frac{(1-4\xi)m^3}{4\pi^2|2x|}K_3(m|2x|)
+
\frac{(2\xi-1)m^2}{8\pi^2x^2}K_2(m|2x|)
\right]
\nonumber
\\
\langle{0}|T_{xx}|0\rangle & = & 0 \ .
\end{eqnarray}
In the massless limit these agree with the literature \cite{bd:book}.
For the massive field $\langle{0}|T_{\mu\nu}|0\rangle$ is non-zero
and has non-vanishing trace for both conformal and minimal coupling. The mass
breaks the conformal invariance of the field.
$\langle{0}|T_{\mu\nu}|0\rangle$ vanishes exponentially as the mass and as $x$ go to infinity,
in contrast to the massless case which behaves as $O(x^{-4})$ for
the minimally coupled field and is identically $0$ for conformal coupling.
It is interesting to note that for conformal coupling the leading order
correction for small mass is $O(m^2)$, while for minimal coupling it is 
$O(m^4)$.

\section{Discussion}

We have calculated the expectation values $\langle{0}|T_{\mu\nu}|0\rangle$
for the massive scalar and Dirac fields on the flat 
spacetimes $M_0$ and $M_-$.
For the scalar field our results on $M_0$ agree with those in
\cite{th:scalar}. For the spinor field on $M_0$ our results are 
twice those of \cite{an:spinor} as expected. On $M_-$ the results for the
massive fields are new. 
Further in the massless limit our expectation values agree
with the previous literature \cite{lm:geon,pl:langlois,bd:em,dw:dewitthi}. 
In all cases the values fall
off exponentially in the large $m$ limit, and the leading order 
correction for small mass is $O(m^2)$.
Further it is noted that for the scalar field in the large $a$ limit on $M_-$
there is an exponential decay in the massive case while for the massless field
the behaviour is $a^{-4}$. 
For the massive field the difference between $\langle{0}|T_{\mu\nu}|0\rangle$
on $M_-$ and the corresponding values on $M_0$ 
vanishes exponentially in the limit of large $x^2+y^2$, while
for the massless field it behaves as $O(r^{-3})$.
As an aside we also presented the expectation values 
$\langle{0}|T_{\mu\nu}|0\rangle$ for a massive scalar 
field on Minkowski space with an infinite straight plane boundary.

While we have focussed the present paper on the stress-energy in its own 
right, our underlying interest in this problem arises from the role of 
$M_0$ and $M_-$ in modelling, in the context of accelerated observers on 
flat spacetimes, the Hawking(-Unruh) effect on respectively the Kruskal 
manifold and the $\mathbb{RP}^3$ geon \cite{lm:geon}. Certain aspects of the thermal 
and non-thermal effects for scalar and spinor fields on $M_0$ and $M_-$ 
are at present understood from the viewpoint of Bogoliubov transformations 
and particle detector analyses \cite{lm:geon,pl:langlois}, but the connections between 
(non-)thermality and stress-energy remain less clear. We view our results, 
in conjunction with those in \cite{lm:geon,pl:langlois}, 
as data points to which we anticipate 
future work on this question to provide a deeper understanding.

\subsection*{Acknowledgements}

I would like to thank Jorma Louko for many useful discussions and
for reading the manuscript. This work was supported by the University of
Nottingham and the States of Guernsey Education Department.

\end{document}